\documentclass[runningheads]{llncs}
\usepackage{graphicx}
\usepackage{color}
\usepackage{tabularx} 
\usepackage{booktabs} 
\newcolumntype{L}[1]{>{\raggedright\arraybackslash}p{#1}}
\newcolumntype{C}[1]{>{\centering\arraybackslash}p{#1}}
\newcolumntype{R}[1]{>{\raggedleft\arraybackslash}p{#1}}

\usepackage{amssymb}  
\usepackage{listings} 
\usepackage{xcolor}

\usepackage{url} 

\definecolor{codegreen}{rgb}{0,0.6,0}
\definecolor{codegray}{rgb}{0.5,0.5,0.5}
\definecolor{codepurple}{rgb}{0.58,0,0.82}
\definecolor{backcolour}{rgb}{0.95,0.95,0.92}

\lstdefinestyle{mystyle}{
    backgroundcolor=\color{backcolour},
    commentstyle=\color{codegreen},
    keywordstyle=\color{magenta},
    numberstyle=\tiny\color{codegray},
    stringstyle=\color{codepurple},
    basicstyle=\ttfamily\footnotesize,
    breakatwhitespace=false,
    breaklines=true,
    captionpos=b,
    keepspaces=true,
    numbers=left,
    numbersep=5pt,
    showspaces=false,
    showstringspaces=false,
    showtabs=false,
    tabsize=2
}
\begin{document}
\title{Performance Assessment of OpenMP Compilers Targeting NVIDIA V100 GPUs}
\titlerunning{Performance Assessment of OpenMP Compilers for V100 GPUs}
%
\author{Joshua Hoke Davis\inst{1}\orcidID{0000-0002-6704-0520} \and
Christopher Daley\inst{2} \and
Swaroop Pophale\inst{3} \and
Thomas Huber\inst{1} \and
Sunita Chandrasekaran\inst{1} \and
Nicholas J. Wright\inst{2}}

\authorrunning{J. Davis et al.}
%
\institute{University of Delaware, Newark DE 19716, USA \and
National Energy Research Scientific Computing Center, Lawrence Berkeley National Laboratory, Berkeley CA 94720, USA \and
Oak Ridge National Laboratory, Oak Ridge TN 37830, USA\\
}
\maketitle              
\begin{abstract}
Heterogeneous systems are becoming increasingly prevalent. In order to exploit the rich compute resources of such systems, robust programming models are needed for application developers to seamlessly migrate legacy code from today's systems to tomorrow's. Over the past decade and more, directives have been established as one of the promising paths to tackle programmatic challenges on emerging systems. This work focuses on applying and demonstrating OpenMP offloading directives on five proxy applications. We observe that the performance varies widely from one compiler to the other; a crucial aspect of our work is reporting best practices to application developers who use OpenMP offloading compilers. While some issues can be worked around by the developer, there are other issues that must be reported to the compiler vendors. By restructuring OpenMP offloading directives, we gain an 18x speedup for the su3 proxy application on NERSC's Cori system when using the Clang compiler, and a 15.7x speedup by switching max reductions to add reductions in the laplace mini-app when using the Cray-llvm compiler on Cori. 

\end{abstract}
\section{Introduction}
\label{sec:intro}
Of the 500 supercomputers on the Top500 list, a full thirty percent (150 systems) use many-core technologies, such as NVIDIA Volta GPUs or Intel Xeon Phi many-core co-processors \cite{top500}. This is up from 133 systems in the list one year ago. Furthermore, seven of the top ten supercomputers on the latest list use many-core technology. Heterogeneous architectures, those which use co-processors or accelerators in addition to a main processor, are valued for their energy efficiency and promise significant performance gains for applications that can make use of them. However, programming for these platforms, and in particular, porting existing applications to these platforms, poses a significant challenge. Scientific programmers look forward to taking advantage of these powerful architectures without having to learn the exact hardware details or make significant changes to their applications, which can often exceed tens of thousands of lines of code.

Numerous programming models and tools exist for programming heterogeneous systems, including CUDA~\cite{cuda}, OpenCL~\cite{opencl}, and Kokkos~\cite{kokkos}. Directive-based models such as OpenACC~\cite{openacc} and OpenMP~\cite{openmp} are popular solutions, as they offer a useful degree of abstraction over various hardware types with a unified interface, and reduce the work needed to accelerate an application, requiring only "hints" or annotations to be added to the compiler. The OpenMP model introduced support for offloading code (with the target directive) to accelerators, co-processors, or many-core processors from version 4.0 (released 2013), and has continued to add and update features through versions 4.5 (released 2015) and 5.0 (released 2018). 

To understand the value of offloading support in OpenMP, we highlight the following points from the 2018 NERSC-10 workload analysis: more than eighty percent of the NERSC community uses OpenMP for parallel programming, making OpenMP far and away the most widely-adopted model in use at NERSC, and fifty-one percent of the NERSC workload is already either fully or partially implemented on GPUs \cite{nerscwa}. Meanwhile, the 2019 OLCF Operational Assessment indicates that its three allocation programs all used more than 75\% of their hours spent on Summit running GPU-enabled jobs, with INCITE reaching 94\% GPU-enabled hours \cite{olcfoa}.

Knowing that heterogeneous architectures are only continuing to grow in popularity, it is critical that users understand the status of the various vendor compilers which support OpenMP offloading. Application developers must be able to make an informed choice of compiler based on which particular offloading features their application uses. Understanding cases in which identical OpenMP directives can show highly variable performance across compilers is essential to making such a decision. And, where compilers exhibit performance differences, understanding the underlying reasons in the implementation for those differences is useful not only for improving the portability of an application between compilers but also for giving specific feedback to vendors about the limitations of their existing implementations.

The main contributions of this work are as follows:
\begin{itemize}
    \item Identify five benchmarks and proxy applications (mini-apps) which characterize the performance of OpenMP offloading features used by major applications and exhibit performance differences across compilers that are of interest to developers.
    \item Quantify performance differences across state-of-the-art compilers for the benchmarks and proxy applications selected.
    \item Explain the observed differences in performance between implementations by using profiling tools and performance metrics, making use of an execution time decomposition methodology, where needed, to quantify the impacts of kernel launch latency and OpenMP runtime overhead.
    \item Make recommendations to application developers regarding the best practices for performance portable OpenMP offloading, guided by insights into the causes of slowdowns in kernels derived from real-world applications.
\end{itemize}
The remainder of this paper is organized as follows. In Section~\ref{sec:relatedwork} we discuss related work in compiler comparison for heterogeneous architectures. In Section~\ref{sec:suitesetup} we describe the five mini-apps selected, as well as the environment and methodology used to test these mini-apps. Section~\ref{sec:results} shows the results for each of the five mini-apps, both in terms of general performance and specific insights gained from profiling, and Section~\ref{sec:discussion} sets out recommendations to application developers based on the insights gained. Finally, in Section~\ref{sec:conclusion} we conclude the paper and identify directions for future work.

\section{Related Work}
\label{sec:relatedwork}
Several existing works narrate the use of OpenMP offloading features for many-core processors and accelerators such as GPUs. These include performance analysis of TeaLeaf and CloverLeaf~\cite{martineau2016evaluating}, as well as LULESH~\cite{bercea2015performance}, which uses OpenMP 4.0. Larrea et al.~\cite{larrea2016early} show preliminary lessons learned writing portable code using OpenMP 4.0. Gayatri et al.~\cite{gayatri2018case} used a material science application with OpenMP 4.5 to compare and contrast with OpenACC, showing that an unchanged OpenMP GPU version of the code was ill-suited for CPU execution. ExaHyPE, an Exascale Hyperbolic PDE design~\cite{weinzierl2020exahype} used a pragma-based GPU parallelization approach for object-oriented code, and documented lessons learned. Several other related works include demonstrating GPU support for OpenMP offloading features in compilers in Flang/Clang~\cite{bertolli2015integrating,ozen2018openmp}, a proof-of-concept implementation of offloading for FPGA based accelerators~\cite{sommer2017openmp,knaust2019openmp}, and an interprocedural statical analysis heuristic at runtime to select optimal grid sizes for offloaded target team constructs~\cite{tiotto2019openmp}, among others. 

There are publicly available benchmark suites to evaluate heterogeneous application performance, e.g SPEC-ACCEL \cite{Juckeland2014,Juckeland2016} and Rodinia \cite{Shuai2009}. The performance of the SPEC-ACCEL benchmark suite was evaluated on multiple platforms using multiple OpenMP offloading and OpenACC compilers by Boehm et al. \cite{Swen2018}. Here, the authors reported a list of compilation/runtime errors for individual benchmarks as well as benchmark execution time, however, there was little detail about reasons for the observed performance with different compilers. The Rodinia benchmark suite was used to evaluate OpenMP offloading Unified Memory performance by Mishra et al. \cite{Mishra2017}. The OpenMP offloading and OpenACC performances of four mini-apps were evaluated across platforms and compilers by Larrea et al. \cite{Larrea2020}. Larrea et al. described the development coding challenges, portability issues and performance, but did not go into detail about the reasons for poor performance reported. A detailed evaluation of the overhead of different OpenMP compilers was performed by Diaz et al. \cite{Diaz2019}, however, this had a narrow focus on the overhead of individual OpenMP constructs.

In contrast to existing related work, this paper focuses on a set of mini-applications, thus forming a suite of codes using two major systems: NERSC Cori and ORNL Summit. We explore the compatibility of the mini-apps with 7 compilers including 5 OpenMP offloading, 1 OpenACC, and 1 CUDA compiler to quantify and document performance differences across compilers \textit{and} offer recommendations to application developers for usability and best practices for OpenMP offloading compilers.


\section{Mini-apps Suite and Experimental Setup}
\label{sec:suitesetup}
\subsection{Mini-apps suite}
\label{subsec:miniapps}
The suite is made up of mini-apps chosen for their focus on offloading kernels, diverse characteristics, and their ability to be compiled by all available compilers. The following benchmarks and proxy applications were selected for this paper:
\begin{enumerate}
    \item \textbf{su3}~\cite{SU3} is a matrix-matrix multiply code using complex numbers. It is extracted from MILC (MIMD Lattice Computation), a Lattice QCD (Quantum Chromodynamics) code.
    \item \textbf{babelStream} \cite{BabelStream} is a memory bandwidth benchmark implemented in multiple programming models. It measures the rate of transfer to and from the global device memory with a number of computational kernels, including dot, add, mul, copy, and triad.
    \item \textbf{laplace} (ported from \cite{chandrasekaran2017openacc}) is an implementation of an iterative Jacobi method Laplace equation solver, which launches multiple small stencil update kernels and uses the OpenMP reduction clause to check for convergence.
    \item \textbf{gpp} \cite{BerkeleyGW-kernels} is a proxy application for the generalized plasmon-pole model from BerkeleyGW, a many-body perturbation theory code. gpp relies on an reduction to compute its final result.
    \item \textbf{ToyPush} (ported from \cite{ToyPush}) is a proxy application 
    for the electron push phase in XGC1, a particle-in-cell simulation code for magnetically-confined fusion plasma. It is similar to laplace in that it launches a large number of short-running kernels.
\end{enumerate}

\subsection{Systems and Compilers}
\label{subsec:environment}
All results shown in this paper use NERSC's Cori machine (GPU testbed nodes) and the Summit supercomputer at the Oak Ridge National Laboratory (ORNL). Table~\ref{tbl:hardware} shows the hardware details of these systems.

\begin{table}
\caption{Overview of the Cori-GPU and Summit systems.}
\centering
\begin{tabular}{@{}L{4cm}L{4cm}L{4cm}@{}}
\toprule
 & \textbf{Cori-GPU} & \textbf{Summit} \\
\midrule
Node architecture & Cray CS-Storm 500NX & IBM AC922 \\
Node CPUs & 2 $\times$ Intel Skylake & 2 $\times$ IBM Power 9 \\
Available cores per CPU & 20 @ 2.40 GHz & 21 @ 3.07 GHz \\
Node GPUs & 8 x 16 GB NVIDIA V100 & 6 x 16 GB NVIDIA V100 \\
CPU-GPU interconnect & PCIe 3.0 switch & NVLink 2.0 \\
\bottomrule
\end{tabular}
\label{tbl:hardware}
\end{table}

Table~\ref{tbl:software} shows the compilers tested for each mini-app, where possible. Because PGI support for OpenMP offloading is still under development, PGI was tested using an OpenACC equivalent implementation of each code. Note that the Clang 11 versions used on Cori are both the same in-development version. The Cray Classic compiler (CCE 9.0.0) refers to the Cray C/C++ compiler that uses proprietary Cray compiler technology, in Cray CCE 10.0.0 the C/C++ compilers have been replaced with Cray enhanced LLVM and clang. This not only means that nearly all of the compiler flags are different, but also that the performance will be different. Table~\ref{tbl:compatibility} shows which of our mini-apps can be compiled and run with which compilers. A status of NI indicates that the mini-app is not implemented in the required programming model for that compiler, while CE and RE indicate compiler and runtime errors, respectively. LLVM's Fortran compiler Flang does not have complete support for OpenMP offloading features, so for the ToyPush application (the sole Fortran app tested) LLVM results cannot be shown.

Throughout this paper, application results are verified whenever the app runs to completion. Each compiler was used with the most aggressive optimization flags enabled, i.e. \verb|-Ofast| (or equivalent if named differently). 

\begin{table}
\caption{Compilers and GPU offloading methods evaluated on the Cori-GPU and Summit systems}
\centering
\begin{tabular}{@{}L{2.5cm}L{2.5cm}L{3.5cm}L{3.4cm}@{}}
\toprule
\textbf{Compiler} & \textbf{GPU offload} & \textbf{Cori-GPU version} & \textbf{Summit version} \\
\midrule
NVCC & CUDA & 10.2.89 & - \\
NVIDIA/PGI & OpenACC & 20.4 & - \\
Cray CCE & OpenMP & 10.0.0 (LLVM version) & - \\
Cray CCE & OpenMP & 9.0.0 (Classic version) & - \\
IBM XL & OpenMP & - & 16.1.1-5 \\
LLVM/Clang & OpenMP & 11.0.0-git (\#17d8334) & 11.0.0-git (\#17d8334) \\
GNU/GCC & OpenMP & - & 9.1.0 \\
\bottomrule
\end{tabular}
\label{tbl:software}
\end{table}

\begin{table}
\caption{Compatibility of mini-apps with each compiler. (NI: No implementation for required programming model; RE: Runtime Error)}
\centering
\begin{tabular}{@{}L{2.7cm}L{1cm}L{1.8cm}L{1.5cm}L{1.2cm}L{1.6cm}L{1.7cm}@{}}
\toprule
\textbf{Compiler} & \textbf{su3} & \textbf{babelStr.} & \textbf{laplace} & \textbf{gpp} & \textbf{ToyPush}  \\
\midrule
NVCC (CUDA) & \checkmark & \checkmark & NI & NI & NI \\
PGI (OpenACC) & \checkmark & \checkmark & \checkmark & \checkmark & \checkmark \\
Cray-llvm & \checkmark & \checkmark & \checkmark & \checkmark & \checkmark \\
Cray-classic & \checkmark & \checkmark & \checkmark & \checkmark & \checkmark \\
XL & \checkmark & \checkmark & \checkmark & \checkmark & \checkmark \\
Clang & \checkmark & \checkmark & \checkmark & \checkmark & - \\
GCC & \checkmark & \checkmark & \checkmark & \checkmark & RE \\
\bottomrule
\end{tabular}
\label{tbl:compatibility}
\end{table}

\subsection{Profiling Methods and Tools}
\label{subsec:methods}


Our approach for understanding performance differences across compilers starts from identifying where such performance differences exist. For each mini-app, tested across all compilers it is compatible with (see Table~\ref{tbl:compatibility}), we first record a metric of performance, which varies depending on the nature of the application.
For example, su3 has a figure of merit of GFLOPs.
For more complex apps such as ToyPush or laplace, execution time was used, while for babelStream, which is memory-bound, we measured memory bandwidth. If the chosen metric for a given application is relatively poor for one compiler compared to others, that indicates this compiler is generating inefficient code.

Knowing which compilers perform poorly for a given application, we use profiling tools to uncover the underlying reasons for such poor performance. The two profiling tools used in this study are nvprof and Nsight Compute, both NVIDIA products. nvprof is a command-line profiler for NVIDIA GPUs, which we use to identify GPU activities and kernels that are most time intensive and collect hardware metrics relating to memory use and instruction counts. Nsight Compute, which has a command-line and graphical component, we use to profile the kernels of an application in-depth. Nsight Compute indicates high-level bottlenecks, creates roofline plots, and features a source analysis view which we use to identify high-latency sections of a kernel in both the original source and generated assembly code. 


The choice of metrics to focus on for a particular application depends on our understanding of what the application is computing, as well as the bottlenecks indicated by Nsight Compute. For example, Nsight Compute tells us babelStream's dot kernel is latency bound when compiled with Clang, with low compute and memory utilization. Knowing this, we use source analysis to identify which source lines have the most latency samples, confirming the impact of the OpenMP reduction specifically. Viewing the SASS (Shader Assembly) alongside the source can provide a deeper understanding of where latency specifically arises, such as constant memory load instructions that appear in some codes compiled with Cray-classic.

In other cases, there is less to be learned from the hardware metrics to gain a deep understanding of a kernel, as the application launches many small kernels rather than a few large kernels. In these cases, we expect kernel launch latency and overhead of the OpenMP runtime to be a major cause of performance degradation. The NVTX (NVIDIA Tools Extension) API provides a set of CPU functions to tag parts of software for GPU profiling. With NVTX, we are able to wrap an OpenMP \texttt{target} region in an NVTX range, so that nvprof will specifically time the region. The general form of this approach is shown in Listing~\ref{lst:nvtx}, using the code structure of the laplace mini-app (see Section~\ref{subsec:laplace}).

\noindent\begin{minipage}{\linewidth}
\begin{lstlisting}[language=C, caption=NVTX range markers, label=lst:nvtx]
id1 = nvtxRangeStartA("launch");
#pragma omp target teams distribute parallel for reduction(..) collapse(2)
for (i = 1; i <= height; ++i) {
    // stencil update...
}
nvtxRangeEnd(id1);
\end{lstlisting}
\end{minipage}

The nvprof profiler then gives us an average duration of the range as well as the average time spent on kernels and data movement in that region. Assuming no overlap, this breakdown is made up of three parts, as shown in Equation~\ref{eqn:time}.

\begin{equation}
\label{eqn:time}
NVTX Range Time = GPU Time + CPU Time + Data Movement Time
\end{equation}

Intuitively, this means that starting with the NVTX range total time reported by nvprof and subtracting the average data movement and GPU kernel time leaves us with the CPU time. This time is accounted for primarily as overhead of the OpenMP runtime.

Note that all profiling and sampling data collected has less than 5\% variation between runs. Profiling overhead varies depending on the tool and configuration. Nsight Compute can show 3x-20x slowdowns, while nvprof \textit{without} metric collection shows a minimal slowdown, around 1.1x to 1.2x from our tests. nvprof with metric collection shows a 1.3x to 7x slowdown. All execution times shown are measured \textit{without} profiling tools.

Section~\ref{sec:results} will show the results collected for each mini-app across compilers, as well as our insights into causes for performance differences taken from performance metrics and profiling.


\section{Results}
\label{sec:results}

\begin{figure}
\includegraphics[width=\textwidth]{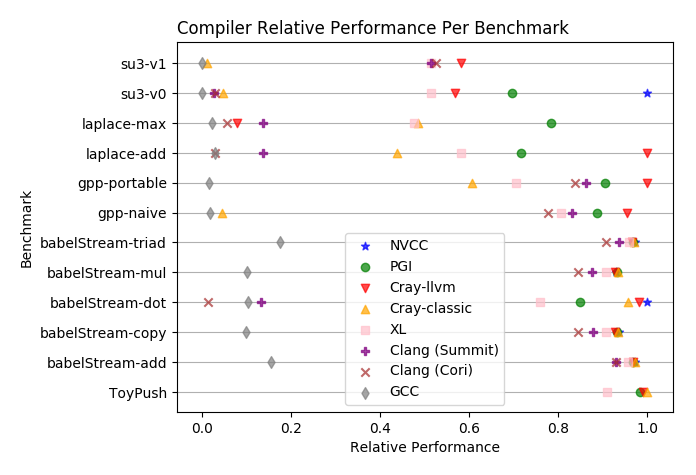}
\caption{Relative performance for each mini-app and compiler.} \label{fig:overview}
\end{figure}

As described in Section~\ref{subsec:methods}, our investigation starts from identifying which applications show drastic performance differences across compilers. Figure~\ref{fig:overview} shows, for each application version and compiler, the degree of variance in performance of the tested compilers. The differences between versions shown for each mini-app will be described in the following subsections. Performance for this figure is on relative scale from zero to one, where one represents the performance of the best-performing compiler, using the most appropriate metric for each benchmark. For example, because babelStream is a memory bandwidth benchmark, the memory bandwidth achieved is used as the performance metric for comparing compiler performance, while laplace uses total execution time. The following subsections will describe the differences between versions for each mini-app, and examine the performance variation in detail.

\subsection{su3}
\label{subsec:su3}
\begin{figure}
\includegraphics[width=\textwidth]{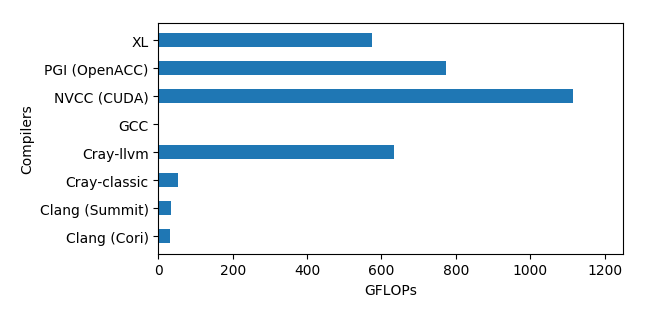}
\caption{GFLOPs per compiler for su3-v0. Performance results are obtained on Cori-GPU, except for "XL", "GCC" and "Clang (Summit)" data points.} \label{fig:su3flops0}
\end{figure}

The su3 mini-app is a matrix-matrix multiply code. Figure~\ref{fig:su3flops0} shows the GFLOPs per compiler, computed using the observed execution time based on the known number of operations the kernel performs. The theoretical peak performance on NVIDIA V100 GPUs based on the calculated arithmetic intensity of the kernel is 1,270 GFLOPs.

Compared to the CUDA baseline, GCC, Clang and Cray-classic stand out as poorly-performing, with $<$1\%, 3\% and 5\% of the CUDA performance, respectively. Using Nsight Compute profiling, we attribute Cray-classic performance to small grid size, and therefore poor device utilization as well as latency issues arising from rapid, intense use of global constant memory. Small grid size is compensated for by increasing the number of teams using the \texttt{num\_teams} clause. By raising the number of teams from 1200, the Clang-tuned value, to 10000, Cray-classic reaches approximately 240 GFLOPs, or a 4.6x speedup.\footnote[1]{This is also a 2.03x speedup compared to the performance of the Cray-classic-chosen default value, which is 81920 teams. Note that Cray-classic ignores \texttt{num\_threads}, as it only considers teams and SIMD parallelism.} In comparison, the best-performing compiler (NVCC) used a grid size of 294912, and the worst-performing compiler (GCC) used a grid size of 1600.

To further investigate Clang performance, we examine DRAM transactions for each compiler. Data is collected using the nvprof command-line profiler. According to the DRAM read and write transaction metrics, su3 performs excess DRAM data movement when compiled with Clang, over 20x more write transactions and 3x more read transactions when compared to the CUDA baseline.

Listing~\ref{lst:su3code0} shows the OpenMP construct arrangement in su3. According to the Clang documentation, this arrangement of directives, specifically, the interleaving for loop between the \texttt{teams} and \texttt{parallel} constructs, causes Clang to choose the non-SPMD mode for code generation. To test if the use of non-SPMD mode is responsible for elevated DRAM transactions, we modify the OpenMP directive structure of su3, as shown in Listing~\ref{lst:su3code1}. This optimized version removes interleaving code between the \texttt{teams} and \texttt{parallel} constructs, and manually distributes the loop iterations between teams.

\begin{lstlisting}[language=C, caption=OpenMP directives in su3-v0, label=lst:su3code0]
#pragma omp target teams distribute 
    for(int i = 0; i < total_sites; ++i) {
#pragma omp parallel for collapse(3)
        // 3 for loops...
\end{lstlisting}

\begin{lstlisting}[language=C, caption=OpenMP directives in su3-v1, label=lst:su3code1]
#pragma omp target teams
#pragma omp parallel
{
    // compute istart, iend for each team ...
    for(int i = istart; i < iend; ++i) {
#pragma omp for collapse(3)
        // 3 for loops...
\end{lstlisting}

\begin{figure}
\includegraphics[width=\textwidth]{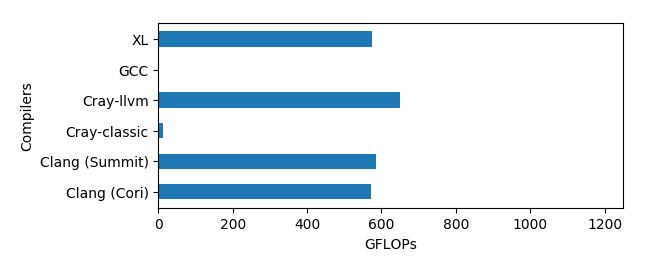}
\caption{GFLOPs per compiler for su3-v1. Performance results are obtained on Cori-GPU, except for "XL", "GCC" and "Clang (Summit)" data points.} \label{fig:su3flops1}
\end{figure}

With su3-v1's modifications, the DRAM transactions for all compilers except Clang remain approximately the same, while Clang DRAM transactions fall to a level matching the other compilers. Examining the GFLOPs per compiler, as shown in Figure~\ref{fig:su3flops1} for su3-v1, shows that this change to the OpenMP directives improves Clang performance substantially, approximately 18x.\footnote[1]{Note that after these modifications, Clang chooses default \texttt{num\_teams} and \texttt{num\_threads} values of 128 and 128, which do not perform as well as our tuned values of 1600 and 64 (4.45x speedup with tuned values compared to defaults).} GCC performance remains 2-3 orders of magnitude worse than all other compilers even with this optimization. The Cray-classic data point did not use our tuned \texttt{num\_teams} value, for comparison purposes.

\subsection{babelStream}
\label{subsec:babelStream}

\begin{figure}
\includegraphics[width=\textwidth]{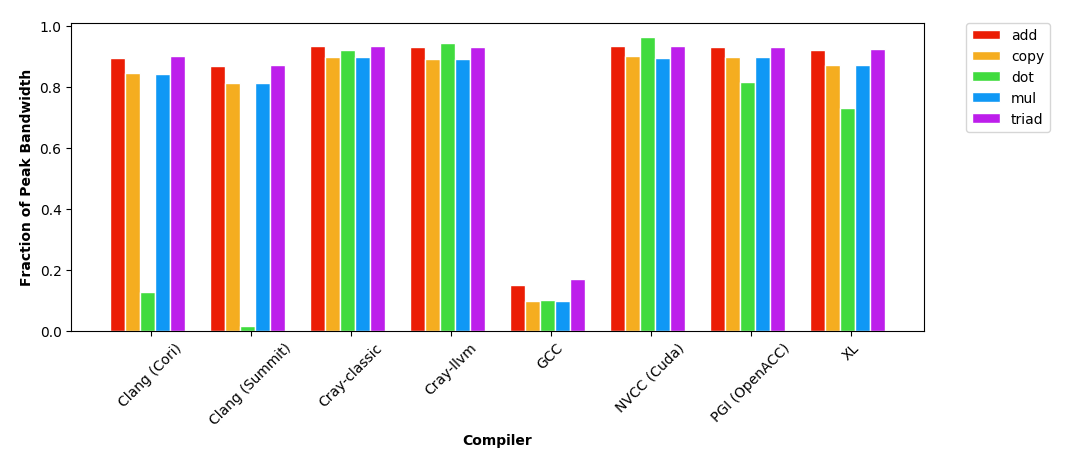}
\caption{Fraction of peak memory bandwidth per compiler for babelStream.} \label{fig:babelbandw}
\end{figure}

The babelStream memory bandwidth benchmark uses a number of simple compute kernels to test memory bandwidth, called dot, copy, add, mul, and triad. The dot kernel is unique in that, unlike the other kernels, it uses a reduction clause in its computation. As babelStream is a global device memory bandwidth benchmark, we expect it to be memory-bound, reaching near-peak memory bandwidth (900 GB/s).
Figure~\ref{fig:babelbandw} shows the measured memory bandwidth for the dot, copy, and add kernels for each compiler.

The dot kernel compiled with Clang stands out in Figure~\ref{fig:babelbandw} as performing poorly. GCC performance for all babelStream kernels is also relatively poor. Nsight Compute identifies the dot kernel, when compiled with Clang, as latency-bound, rather than bandwidth-bound as expected. Nsight Compute Warp State Analysis points out that the kernel has stall issues, with each warp on average spending 28.7 cycles waiting on a barrier, and Source view shows that there are a large number of barrier latency samples collected on the OpenMP directive that has the reduction clause. Taking into account the lack of similar latency issues on any other babelStream kernel, we infer that these barrier samples must arise due to the introduction of the reduction clause.

\subsection{laplace}
\label{subsec:laplace}

\begin{figure}
\includegraphics[width=\textwidth]{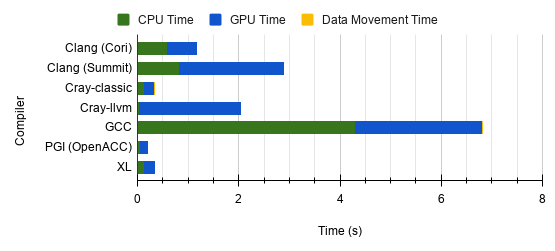}
\caption{Breakdown of execution time in laplace. Results are shown in seconds.} \label{fig:laplacemaxtime}
\end{figure}

The laplace mini-app has two features which we suspect to be possible performance impediments: first, it uses a reduction clause to determine if the computation has converged, and second, it executes a large number of short-running kernels, which would increase the impact of the OpenMP runtime overhead and of any kernel launch latency. As described in Section~\ref{subsec:methods}, NVTX range markers allow us to measure the composition of execution time for a given offloading region, following Equation~\ref{eqn:time}. Figure~\ref{fig:laplacemaxtime} shows the results of this approach for each compiler.

Cray-llvm performs poorly, due to high GPU time, while Clang and GCC also perform poorly, due to high CPU and GPU time. This high CPU time in Clang indicates a high overhead of the OpenMP runtime, which negatively impacts performance. Clang GPU time is longer on Summit compared to on Cori, which according to Nsight Compute profiling, is caused by elevated (about 10x higher) barrier latency on the max reduction for Clang on Summit, specifically on one move instruction compared to Clang on Cori. This is causing far more warp stalling and thus lower compute and memory utilization.

The most significant limiter of laplace performance is the use of a max \texttt{reduction} clause. Add reductions are shown in the babelStream study to be a source of latency issues in Clang, but did not pose a problem for Cray-llvm. To confirm that the max reduction specifically causes the high Cray-llvm GPU Time seen in Figure~\ref{fig:laplacemaxtime}, we create a version of the laplace app that uses an add reduction rather than a max reduction. Rather than detecting convergence, it merely iterates a fixed number of times, the number of iterations the max reduction version needed to converge. Most compilers show little performance difference between the max and add reduction versions, but Cray-llvm shows a significant difference, a 15.7x speedup in GPU Time using add reduction version.

Profiling with Nsight Compute shows the reasons for this extreme difference in max and add reduction performance in Cray-llvm. Source analysis indicates that the max reduction clause has a large number of Long Scoreboard latency samples. Each warp of the kernel spends on average 53.6 cycles stalled waiting for a L1TEX (global memory) operation, meaning that the kernel is using global memory heavily. Further investigation into the assembly generated for the reduction shows the source: an atomic operation on global memory. By comparison, the add reduction in Cray-llvm has far fewer latency samples, and no similar atomic operation.

The difference between the max and add reduction implementations in Cray-llvm can be further confirmed with hardware metrics. nvprof profiling finds that the metrics \texttt{atomic\_transactions} and \texttt{l2\_atomic\_transactions} are both approximately 2740 times higher for the max reduction version compared to the add reduction version.

Nsight Compute SASS view shows specifically that Cray-llvm uses fewer hardware atomic-add instructions, implying the use of tuned reduction algorithms, e.g. 5-stage hierarchical shuffle-based algorithms. This is not the case for the Clang compiler, which uses a relatively large number of general purpose compare-and-swap atomic instructions. Detailed analysis of the Clang compiler shows a similar count of compare-and-swap atomic instructions for the max and add reductions, implying reuse of the compiler code. The Cray-llvm compiler uses 4 orders of magnitude more atomic instructions than Cray-classic implying use of a general purpose slower code path. Only three compilers, PGI, Cray-classic and XL, generate an efficient max reduction according to Figure~\ref{fig:laplacemaxtime}.



\subsection{gpp}
\label{subsec:gpp}

gpp is a larger mini-app, which uses an add reduction to compute its final result. There are two versions tested for gpp: gpp-portable, which includes the default reduction reconfiguration described below, and gpp-naive, which removes that reconfiguration. Measuring execution time of gpp-portable across compilers we observe consistency across compilers, save for GCC. Examining the use of the \texttt{reduction} clause in gpp-portable, we see a reconfiguration approach to mitigate the impact of reduction slowdowns in some cases, which explains this generally consistent good performance. To understand the possible benefits of this approach, consider how the reduction would usually be done (i.e., as it is in gpp-naive), shown in a simplified form in Listing~\ref{lst:gppnaive}.

\noindent\begin{minipage}{\linewidth}
\begin{lstlisting}[language=C, caption=gpp-naive reduction usage, label=lst:gppnaive]
#pragma omp target teams distribute parallel for simd collapse(2) reduction(+:sum)
    for(/* iterate over ngpown */) {
        for(/* iterate over num_bands */) {
            for(/* interate over ncouls */) {
                // compute values...
                sum += computed_value;
            }
        }
    }
\end{lstlisting}
\end{minipage}

The reduction operator in gpp-naive is placed in the innermost loop, as per usual, so that every iteration of the innermost loop adds to the \texttt{reduction} variable. By comparison, gpp-portable moves the reduction operations one loop up, after the innermost loop inside the middle loop. The innermost loop instead sequentially stores the results of the innermost loop in a local variable, which is only reduced after the inner loop is complete. Listing~\ref{lst:gppport} demonstrates the approach, simplified.
\begin{lstlisting}[language=C, caption=gpp-portable reduction usage, label=lst:gppport]
#pragma omp target teams distribute parallel for simd collapse(2) reduction(+:sum)
    for(/* iterate over ngpown */) {
        for(/* iterate over num_bands */) {
            double local_sum = 0.0;
            for(/* interate over ncouls */) {
                // compute values...
                local_sum += computed_value
            }
            sum += local_sum;
        }
    }
\end{lstlisting}

\begin{figure}
\includegraphics[width=\textwidth]{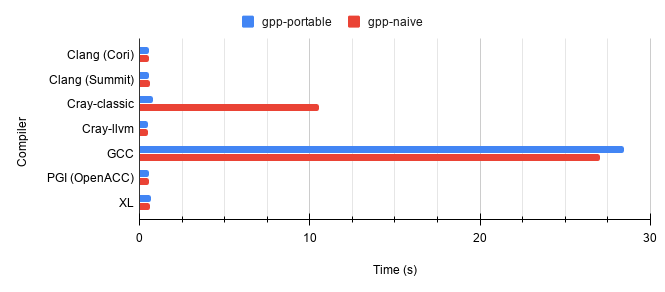}
\caption{Comparing execution time of gpp-portable and gpp-naive.} \label{fig:gpp}
\end{figure}

Figure~\ref{fig:gpp} compares the execution time of gpp-portable and gpp-naive, and as expected, gpp-naive is generally slower than gpp-portable. gpp-portable, compared to gpp-naive, shows a 1.02x to 1.05x speedup in kernel time for the Clang, Cray-llvm, and PGI compilers, a 13.4x speedup for Cray-classic, and a 0.88x and 0.95x slowdown for XL and GCC (meaning the change harms GCC and XL performance, and only slightly improves performance for other compilers except Cray-classic). GCC is also relatively poorly-performing compared to other compilers. The particularly poor Cray-classic performance on gpp-naive can be attributed to elevated device memory activity, as it shows approximately 3714 times more bytes transferred to and from device memory compared to Clang's version of gpp-naive.

\subsection{ToyPush}
\label{subsec:toypush}

\begin{figure}
\includegraphics[width=\textwidth]{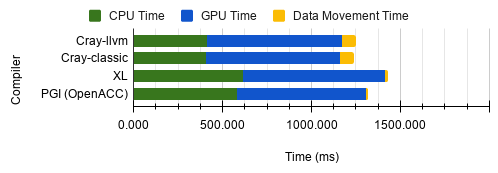}
\caption{Breakdown of execution time in ToyPush. Results are shown in milliseconds.} \label{fig:toypush}
\end{figure}

ToyPush provides an example of a larger mini-app, taken from a real-world application, that exemplifies the pattern shown in laplace. Like laplace, ToyPush launches a large number of short-running kernels. Figure~\ref{fig:toypush} shows the results, using the NVTX range technique shown in Section~\ref{subsec:methods}. The time shown in this figure is the total time spent on each activity type within the OpenMP offloading region, as demarcated in Listing~\ref{lst:nvtx}.

The relatively elevated CPU Time for XL and PGI corresponds with the total time taken to execute the mini-app, as XL and PGI were the two poorest-performing compilers by that metric. While Clang was shown in the laplace analysis to have the highest OpenMP runtime overhead, the Flang (LLVM Fortran) compiler is not able to compile this OpenMP offloading code and was thus not used in our study. Even so, Figure~\ref{fig:toypush} confirms that elevated runtime overhead impacts mini-apps derived from real-world applications, and if the future Flang compiler has a similar high overhead as in Clang, ToyPush performance would be expected to be poor.

Unlike laplace, total GPU time in ToyPush is consistent across compilers. However, note that data movement time for both Cray compilers appears elevated. nvprof profiling indicates that while the Cray compilers use pageable memory, XL and PGI used pinned memory. From discussion with compiler engineers, this is evidence that XL and PGI are performing a memory optimization, copying data to pinned memory before transfer to the GPU, broken into chunks sized to fit into pinned memory.



\section{Discussion}
\label{sec:discussion}

In this section, we summarize and discuss the results. The results in Figure \ref{fig:overview} show that benchmark performance can sometimes vary by up to an order of magnitude across compilers. In Section \ref{sec:discussionA} we will outline the main reasons for these performance differences and in Section \ref{sec:discussionB} we will give guidelines to help application developers achieve higher performance across today's OpenMP offloading compilers.

\subsection{Performance issues across compilers}
\label{sec:discussionA}

Three mini-apps/kernels were impacted by poor OpenMP reduction performance: laplace, gpp, babelStream-dot. We find that all the compilers generate an efficient OpenMP add reduction except for LLVM/Clang, which uses a relatively large number of general purpose compare-and-swap atomic instructions to implement an OpenMP reduction. We also find that only PGI, Cray-classic and XL compilers generate an efficient max reduction. The Cray-llvm implementation uses 4 orders of magnitude more atomic instructions than Cray-classic implying the use of a general purpose slower code path. The LLVM/Clang compiler show a similar count of compare-and-swap atomic instructions for the max and add reductions implying reuse of the compiler code. It is likely that Cray-llvm performance will improve over time as HPE incorporates more of the optimizations from Cray-classic into Cray-llvm. We hope to see OpenMP reduction performance become a priority optimization in the open-source LLVM/Clang compiler. This is because many applications, including laplace and gpp, benefit from high performance reductions. We also note that the Clang compiler, the only compiler we test on both Cori and Summit, generally shows similar performance between the two platforms.

The su3-v0 and gpp-naive mini-apps are impacted by unexpected data movement between GPU device memory and GPU registers. The su3-v0 mini-app is characterized by \texttt{teams distribute} and \texttt{parallel for} directives on separate loops. Our results show that the Clang compiler generated approximately 20x more data movement than the other tested compilers. The excess data movement is due to the LLVM/Clang compiler using a general purpose code generation path when OpenMP directives are split in this way. It is needed because it is unknown whether each GPU thread will execute identical code on independent data. Given the uncertainty, the compiler flushes memory to ensure a consistent view of memory between successive parallel regions as well as between the team master and the parallel threads. The other compilers do not generate excess data movement for su3-v0 because of compiler optimization passes, e.g. the XL compiler uses interprocedural static compiler analysis to determine that all threads in a team execute the same code \cite{tiotto2019openmp}. The second mini-app, gpp-naive, sums data contributions over 4 nested loops. We find that Cray-classic generated code has 4 orders of magnitude more device data movement than the corresponding Clang generated code. The data movement significantly decreases in the gpp-portable version of the mini-app, where the programmer uses extra private variables to do a per-thread sum over the inner two loops before adding this sum to the OpenMP reduction variable. This indicates that it is sometimes necessary to manually exploit data reuse rather than relying on the compiler.

Finally, the Laplace mini-app is impacted by OpenMP runtime overhead for the LLVM/Clang compiler. This mini-app uses a small problem size that makes it sensitive to target region latency. The surprising observation is that target region latency can be significantly larger than kernel launch latency. In the absence of reductions, we measure $50\mu s$ target region time for the LLVM/Clang compiler and $7-20\mu s$ target region time for the proprietary compilers. This indicates that the management of an OpenMP device data environment is particularly high for the LLVM/Clang compiler. CPU profiling of this overhead to determine its cause is an area for future study.

\subsection{Recommendations to programmers}
\label{sec:discussionB}

The results in the paper draw attention to some mini-apps/kernels which perform relatively well across all tested compilers. The babelStream-triad/\-mul/\-copy/\-add kernels and ToyPush mini-app perform consistently well. If we expand the list to include mini-apps with a median performance of $>0.8$ relative to the best performing compiler then it also includes babelStream-dot, gpp-portable and gpp-naive. The characteristics of these applications include
\begin{itemize}
    \item Minimal data movement between CPU and GPU
    \item Combined \texttt{teams distribute parallel for} constructs
    \item Minimal use of OpenMP reductions
    \item Average GPU kernel runtime $>50 \mu s$
\end{itemize}

None of the mini-apps spent a large fraction of time moving data between CPU and GPU, however, this is often the biggest bottleneck in newly ported applications. We found non-negligible time spent in ToyPush and identified an interesting optimization in the PGI and XL compilers where pinned memory was used to efficiently transfer data between CPU and GPU. This could be important for users who have applications more bound by CPU-GPU data movement time than the mini-apps in our sample.

We recommend that the combined \texttt{teams distribute parallel for} constructs are used where possible. In cases where this is not possible, we draw attention to our experience with su3-v0, which has \texttt{teams distribute} and \texttt{parallel for} on separate loops. We found two different reasons for poor performance with the Cray-classic and LLVM/Clang compiler. The Cray-classic compiler performs poorly because the compiler selected a poor kernel launch configuration; we were able to improve performance by manually increasing the number of teams. The LLVM/Clang compiler performs poorly because the compiler used a general purpose code generation mode which resulted in more memory flushes.

We recommend that OpenMP reductions are used only where necessary because we found mixed performance across compilers. In addition, max reductions sometimes perform significantly worse than add reductions. One suggested method to improve reduction performance in Clang is to add the option \verb|-fopenmp-cuda-teams-reduction-recs-num=<num>|, with \verb|<num>| set to the number of loop iterations. However, in our experience this never led to more than a 10\% speedup with any mini-app. Our experience with gpp-portable and gpp-naive is inconclusive about whether it is beneficial for the programmer to use additional private variables to reduce thread local contributions before reducing into OpenMP reduction variables. Using this technique we see significant performance improvement for the Cray-classic compiler but results in a slow down for the XL compiler. In general, we hope that compiler developers prioritize the performance of OpenMP reductions because it is a frequently used parallel pattern.

We also want to make programmers aware of OpenMP runtime overheads. We found that Laplace mini-app performance is very sensitive to OpenMP target region latency. The overhead is highest for the GCC and LLVM/Clang compiler. It should be noted that OpenMP provides a huge convenience to the programmer by enabling a single variable name to refer to data in a host and a device data environment. The cost of this convenience is that launch latencies are higher than using CUDA or some other lower-level API. Therefore programmers must make sure to send sufficient work to the GPU to justify the data transfer overhead. 

We specifically recommend that, at this time, programmers use the GCC compiler primarily for correctness, not for performance. GCC was consistently a low-end outlier in our study. We hope this is temporary as OLCF is working with Mentor Graphics to improve the performance of OpenMP 5.0 features while specifically focusing on GPU offloading directives.

It is common for many developers to use roofline analysis to evaluate performance on CPU+GPU systems. We suggest that developers supplement this analysis with some additional measurements based on our experience with OpenMP mini-apps. These measurements are DRAM read/write transactions, average kernel runtime, and atomic instructions for at least 2 compilers. We have seen that DRAM read and write transactions can sometimes be much higher for some compilers than for others. If one were to rely on roofline analysis only, then it can seem like the mini-apps are achieving close to the memory bandwidth roofline, even though this would just be artifact of excessive data movement. We suggest that average kernel runtime should be measured to identify cases where runtime is less than $<50 \mu s$. Some compilers, e.g. LLVM/Clang, would likely spend at least this much time in just CPU runtime overhead. In this case, developers should look for opportunities to fuse target region code or investigate opportunities to launch target regions asynchronously using nowait or through multiple host threads. Finally, we suggest measuring atomic instruction latency to understand if a compiler OpenMP reduction implementation is the reason for poor OpenMP reduction performance.


\section{Conclusions and Future Work}
\label{sec:conclusion}

This paper fills a gap in the literature, comparing OpenMP offloading compilers with multiple mini-applications derived from real-world apps and most importantly examining performance differences in detail, uncovering the specific causes for slowdowns in implementations. We do so in a way that allows us to make specific and useful recommendations to application developers. These recommendations should allow developers to avoid common performance snags found in the available compilers. Broadly, our findings regarding compiler performance show that runtime overhead in compilers tends to have a bigger negative impact on performance than architecture-specific vendor optimizations have a benefit. Without changes in the compilers, these overheads will not go away when moving to the next generation of GPUs or accelerators; the issues are more fundamental to the compilers.

Future work in this area includes studying additional GPUs, such as those produced by Intel and AMD, as well as NVIDIA Ampere (to be used in the upcoming Perlmutter system). With the addition of these hardware platforms, there is the opportunity to test additional compilers, such as Intel ICC and AMD AOMP. Further, the PGI compiler will support OpenMP offloading in the future. Beyond merely expanding coverage, a long-term goal would be creating a suite similar to the OpenMP Verification and Validation project \cite{Diaz2019}, a publicly-available, well-documented suite, with a comprehensive set of kernel-only mini-apps extracted from real applications. This would provide application and compiler developers a tool for understanding the performance strengths and weaknesses of the available compilers, on various architectures, with open source and transparent tests that anyone can run.


\section*{Acknowledgements}
This research used resources of the National Energy Research Scientific Computing Center (NERSC), a U.S. Department of Energy Office of Science User Facility operated under Contract No. DE-AC02-05CH11231. This research also used resources of the Oak Ridge Leadership Computing Facility, which is a DOE Office of Science User Facility supported under Contract DE-AC05-00OR22725. The authors would like to thank Doug Doerfler and Rahul Gayatri for helpful discussion about the su3 benchmark and useful research directions for this project.

\bibliographystyle{splncs04}
\bibliography{references}

\begin{thebibliography}{10}
\providecommand{\url}[1]{\texttt{#1}}
\providecommand{\urlprefix}{URL }
\providecommand{\doi}[1]{https://doi.org/#1}

\bibitem{nerscwa}
Austin, B.: Nersc-10 workload analysis (data from 2018) (2020),
  \url{https://portal.nersc.gov/project/m888/nersc10/workload/N10_Workload_Analysis.latest.pdf}

\bibitem{bercea2015performance}
Bercea, G.T., Bertolli, C., Antao, S.F., Jacob, A.C., Eichenberger, A.E., Chen,
  T., Sura, Z., Sung, H., Rokos, G., Appelhans, D., et~al.: Performance
  analysis of openmp on a gpu using a coral proxy application. In: Proceedings
  of the 6th International Workshop on Performance Modeling, Benchmarking, and
  Simulation of High Performance Computing Systems. pp. 1--11 (2015)

\bibitem{bertolli2015integrating}
Bertolli, C., Antao, S.F., Bercea, G.T., Jacob, A.C., Eichenberger, A.E., Chen,
  T., Sura, Z., Sung, H., Rokos, G., Appelhans, D., et~al.: Integrating gpu
  support for openmp offloading directives into clang. In: Proceedings of the
  Second Workshop on the LLVM Compiler Infrastructure in HPC. pp. 1--11 (2015)

\bibitem{Swen2018}
Boehm, S., Pophale, S., Vergara~Larrea, V.G., Hernandez, O.: Evaluating
  performance portability of accelerator programming models using spec accel
  1.2 benchmarks. In: Yokota, R., Weiland, M., Shalf, J., Alam, S. (eds.) High
  Performance Computing. pp. 711--723. Springer International Publishing, Cham
  (2018)

\bibitem{chandrasekaran2017openacc}
Chandrasekaran, S., Juckeland, G.: OpenACC for Programmers: Concepts and
  Strategies. Addison-Wesley Professional (2017)

\bibitem{Shuai2009}
Che, S., Boyer, M., Meng, J., Tarjan, D., Sheaffer, J.W., Lee, S.H., Skadron,
  K.: Rodinia: A benchmark suite for heterogeneous computing. In: 2009 IEEE
  international symposium on workload characterization (IISWC). pp. 44--54.
  Ieee (2009)

\bibitem{BabelStream}
Deakin, T.: {BabelStream} (2020), \url{https://github.com/UoB-HPC/BabelStream}

\bibitem{SU3}
Doerfler, D.: {su3\_bench} (2020),
  \url{https://gitlab.com/NERSC/nersc-proxies/su3_bench}

\bibitem{BerkeleyGW-kernels}
Gayatri, R.: {BerkeleyGW-kernels} (2020),
  \url{https://gitlab.com/NERSC/nersc-proxies/BerkeleyGW-Kernels-CPP}

\bibitem{gayatri2018case}
Gayatri, R., Yang, C., Kurth, T., Deslippe, J.: A case study for performance
  portability using openmp 4.5. In: International Workshop on Accelerator
  Programming Using Directives. pp. 75--95. Springer (2018)

\bibitem{Juckeland2014}
Juckeland, G., Brantley, W., Chandrasekaran, S., Chapman, B., Che, S.,
  Colgrove, M., Feng, H., Grund, A., Henschel, R., Hwu, W.M.W., Li, H.,
  M{\"u}ller, M.S., Nagel, W.E., Perminov, M., Shelepugin, P., Skadron, K.,
  Stratton, J., Titov, A., Wang, K., van Waveren, M., Whitney, B., Wienke, S.,
  Xu, R., Kumaran, K.: Spec accel: A standard application suite for measuring
  hardware accelerator performance. In: Jarvis, S.A., Wright, S.A., Hammond,
  S.D. (eds.) High Performance Computing Systems. Performance Modeling,
  Benchmarking, and Simulation. pp. 46--67. Springer International Publishing,
  Cham (2015)

\bibitem{Juckeland2016}
Juckeland, G., Hernandez, O., Jacob, A.C., Neilson, D., Larrea, V.G.V., Wienke,
  S., Bobyr, A., Brantley, W.C., Chandrasekaran, S., Colgrove, M., Grund, A.,
  Henschel, R., Joubert, W., M{\"u}ller, M.S., Raddatz, D., Shelepugin, P.,
  Whitney, B., Wang, B., Kumaran, K.: From describing to prescribing
  parallelism: Translating the spec accel openacc suite to openmp target
  directives. In: Taufer, M., Mohr, B., Kunkel, J.M. (eds.) High Performance
  Computing. pp. 470--488. Springer International Publishing, Cham (2016)

\bibitem{opencl}
Khronos: Opencl (2020), \url{https://www.khronos.org/opencl/}

\bibitem{knaust2019openmp}
Knaust, M., Mayer, F., Steinke, T.: Openmp to fpga offloading prototype using
  opencl sdk. In: 2019 IEEE International Parallel and Distributed Processing
  Symposium Workshops (IPDPSW). pp. 387--390. IEEE (2019)

\bibitem{kokkos}
Kokkos: kokkos/kokkos (2020), \url{https://github.com/kokkos/kokkos}

\bibitem{ToyPush}
Koskela, T.: {ToyPush} (2017), \url{https://github.com/tkoskela/toypush}

\bibitem{larrea2016early}
Larrea, V.V., Joubert, W., Lopez, M.G., Hernandez, O.: Early experiences
  writing performance portable openmp 4 codes. In: Proc. Cray User Group
  Meeting, London, England (2016)

\bibitem{martineau2016evaluating}
Martineau, M., McIntosh-Smith, S., Gaudin, W.: Evaluating openmp 4.0's
  effectiveness as a heterogeneous parallel programming model. In: 2016 IEEE
  International Parallel and Distributed Processing Symposium Workshops
  (IPDPSW). pp. 338--347. IEEE (2016)

\bibitem{Mishra2017}
Mishra, A., Li, L., Kong, M., Finkel, H., Chapman, B.: Benchmarking and
  evaluating unified memory for openmp gpu offloading. In: Proceedings of the
  Fourth Workshop on the LLVM Compiler Infrastructure in HPC. LLVM-HPC’17,
  Association for Computing Machinery, New York, NY, USA (2017).
  \doi{10.1145/3148173.3148184}, \url{https://doi.org/10.1145/3148173.3148184}

\bibitem{Diaz2019}
Monsalve~Diaz, J.M., Friedline, K., Pophale, S., Hernandez, O., Bernholdt, D.,
  Chandrasekaran, S.: Analysis of openmp 4.5 offloading in implementations:
  Correctness and overhead. Parallel Computing  \textbf{89},  102546 (08 2019).
  \doi{10.1016/j.parco.2019.102546}

\bibitem{cuda}
NVIDIA: About cuda (2020), \url{https://developer.nvidia.com/about-cuda}

\bibitem{olcfoa}
OLCF: Operational assessment 2019 oak ridge leadership computing facility
  (2020),
  \url{https://www.olcf.ornl.gov/wp-content/uploads/2020/06/2019OLCF_OAR_FINAL.pdf}

\bibitem{openacc}
OpenACC: About openacc (2020), \url{https://www.openacc.org/about}

\bibitem{openmp}
OpenMP: Openmp specifications (2020),
  \url{https://www.openmp.org/specifications/}

\bibitem{ozen2018openmp}
{\"O}zen, G., Atzeni, S., Wolfe, M., Southwell, A., Klimowicz, G.: Openmp gpu
  offload in flang and llvm. In: 2018 IEEE/ACM 5th Workshop on the LLVM
  Compiler Infrastructure in HPC (LLVM-HPC). pp.~1--9. IEEE (2018)

\bibitem{sommer2017openmp}
Sommer, L., Korinth, J., Koch, A.: Openmp device offloading to fpga
  accelerators. In: 2017 IEEE 28th International Conference on
  Application-specific Systems, Architectures and Processors (ASAP). pp.
  201--205. IEEE (2017)

\bibitem{tiotto2019openmp}
Tiotto, E., Mahjour, B., Tsang, W., Xue, X., Islam, T., Chen, W.: Openmp 4.5
  compiler optimization for gpu offloading. IBM Journal of Research and
  Development  \textbf{64}(3/4),  14--1 (2019)

\bibitem{top500}
TOP500.org: June 2020 top500 (2020),
  \url{https://www.top500.org/lists/top500/2020/06/}

\bibitem{Larrea2020}
Vergara~Larrea, V.G., Budiardja, R.D., Gayatri, R., Daley, C., Hernandez, O.,
  Joubert, W.: Experiences in porting mini-applications to openacc and openmp
  on heterogeneous systems [published online ahead of print (24 april 2020)].
  Concurrency and Computation: Practice and Experience p. e5780 (2020).
  \doi{10.1002/cpe.5780},
  \url{https://onlinelibrary.wiley.com/doi/abs/10.1002/cpe.5780}

\bibitem{weinzierl2020exahype}
Weinzierl, T.: Exahype’s openmp gpgpu port--lessons learned (2020),
  \url{www.peano-framework.org/wp-content/uploads/2020/08/GPGPUs_Lessons_Learned.pdf},
  self-published

\end{thebibliography}

\end{document}